\documentstyle[preprint,aps,epsfig]{revtex}

\def\be{\begin{equation}}  
\def\ee{\end{equation}}
\def\tsigma{{\mathord{\buildrel{\lower3pt\hbox{$\scriptscriptstyle\leftrightarrow$}}\over \sigma}}}
\def\trho{{\mathord{\buildrel{\lower3pt\hbox{$\scriptscriptstyle\leftrightarrow$}}\over \rho}}}
\def\tL{{\mathord{\buildrel{\lower3pt\hbox{$\scriptscriptstyle\leftrightarrow$}}\over L}}}
\def\tJ{{\bf J}}
\def\tE{{\bf E}}

\def\tI{{\mathord{\buildrel{\lower3pt\hbox{$\scriptscriptstyle\leftrightarrow$}}\over I}}}
\def\tb{{\mathord{\buildrel{\lower3pt\hbox{$\scriptscriptstyle\leftrightarrow$}}\over b}}}
\def\tnu{{\mathord{\buildrel{\lower3pt\hbox{$\scriptscriptstyle\leftrightarrow$}}\over \nu}}}

\begin{document}
\draft 

\title{Theory of  electrical conductivities of  ferrogels  }
\author{J. P. Huang}
\address{
Max Planck Institute for Polymer Research, Ackermannweg 10, 55128, Mainz, Germany, and \\
Department of Physics, The Chinese University of Hong Kong, Shatin, NT, Hong Kong
}
\maketitle

\begin{abstract}

Conductive organic polymers can be formulated with  polymers that incorporate  fine dispersed metallic particles.
 In this work, we present a general model for  ferrogels  which are  chemically cross-linked polymer networks swollen with a ferrofluid.  Our aim is to study the effect of the shape and/or material (conductivity) anisotropy on the effective electrical  conductivity of the ferrogel in the presence  of an external magnetic field.
Our theory can reproduce the known results, and provides a link between the particle property and  orientation distribution and the effective   electrical  conductivity.  To this end, we find that material (conductivity) anisotropies are more important to yield a high effective electrical  conductivity than shape anisotropies, while magnetic fields can offer a  correction.

\end{abstract}



\newpage

\section{introduction}

Conductive polymers~\cite{Science1} have received much attention due to its extensive industrial applications (for a review, see Ref.~\cite{Book1}), such as in nonlinear optical waveguides, vapor-phase detectors, twisted nematic liquid crystal displays, and so on. 
In fact, the conductive organic polymers~\cite{Science1} can be formulated with  polymers that incorporate conductive additives such as carbon black~\cite{Book1} or fine dispersed metallic particles~\cite{XRef7}. This is because the presence of certain additives in the polymer can enhance the effective electrical conductivity.
In particular, recently organic polymer composites filled with conductive metallic particles  received much attention in scientific research~\cite{JPCB0,JPCB1,JPCB3}. For such polymer composites, their  electrical characteristic  are close to that of metals, whereas the mechanical property and processing method are typical for plastics~\cite{XRef7,JPCB0,JPCB1,JPCB3}.
The effective electrical  conductivity of metal-polymer composite should depend on the  conductivity of particles, the particle shape, the volume fraction of particles, and the spatial  orientation or alignment  of particles in the composite.

Ferrogels~\cite{P1,P2,P3,Pleiner} are a new class of magneto-controlled elastic materials, which are chemically cross-linked polymer networks swollen with a ferrofluid.  A ferrofluid~\cite{ferro} is a colloidal dispersion of monodomain ferromagnetic particles. In the ferrogel, the finely distributed magnetic particles are located in the swelling liquid and attached to the flexible network chains by adhesive forces.  In the absence of an applied field the magnetic moments are randomly oriented, and thus the gel has no net magnetization. If an external field is applied, the magnetic moments tend to align with the field, thus yielding a bulk magnetic moment. In case of ordinary field strengths, the tendency of the dipole moments to align with the applied field is partially overcome by thermal agitation. As the strength of field increases, all the particles eventually align their moments along the direction of the field, and as a result, the magnetization saturates. If the field is turned off, the magnetic dipole moments quickly randomize and thus the bulk magnetization is again reduced to zero. In a zero magnetic field a ferrogel presents a mechanical behavior very close to that of a swollen network filled with non-magnetic colloidal particles. In uniform magnetic fields a ferrogel experiences no net force, and hence no  macroscopic shape changes and motion are observed.   Ferrogels have much application ranging from  soft actuators, micromanipulators, and artificial muscles~\cite{P1}, to cancer therapy~\cite{P2}  or as an apparatus for immunoblotting~\cite{P3}.

Anisotropy is a common phenomenon in most materials, and it can be an intrinsic material property or induced by the application of fields. Also, in real applications, the shape of particles may deviate from a perfect spherical shape during fabrication.  In ferrogels, due to the anisotropic particle shape and/or conductivity of the  ferromagnetic particles, the specific spatial orientation of the particles can be realized by using an external magnetic field. Thus, the electric properties of ferrogels can be controlled by reorientation of the particles in an applied magnetic field.


In this paper, we shall focus on an anisotropic ferrogel in which both the particle shape and conductivity possess a tensorial form. Our aim is to study the effects of the shape and/or material (conductivity) anisotropy on the effective electrical  conductivity of the ferrogel, by taking into account the spatial orientation of the particles, which can be affected by the external magnetic field. To this end, we find both the shape anisotropy and the external magnetic field can cause the effective electrical  conductivity to increase. Furthermore, the material (conductivity) anisotropy can increase the effective electrical  conductivity significantly.

This paper is organized as follows. In Sec.~II, we put forth a general model for a ferrogel containing ellipsoidal ferromagnetic particles with anisotropic conductivity. In Sec.~III, we numerically calculate two cases, namely, spherical particles with anisotropic conductivity, and spheroidal particles with isotropic conductivity, in an attempt to focus on the influence of the shape and material anisotropy, respectively. This paper ends with a discussion and conclusion in Sec.~IV.


\section{A general model for Ellipsoidal particles with anisotropic conductivities}

Let us start by considering an ellipsoidal particle with depolarization factors $L_x,$ $L_y,$  $L_z$ along $x,$   $y,$   $z$ axes, respectively. Namely, the depolarization factor has a tensor form
\be
\tL = \left( \begin{array}{ccc}
          L_x & 0 & 0 \\
          0 & L_y & 0 \\
          0 & 0 & L_z
          \end{array}\right).
\ee
The three components satisfy a sum rule, $L_x+L_y+L_z = 1$~\cite{Landau84}. In the case of  crystalline anisotropic, the conductivity of the particle has a tensor form as
\begin{equation}
\tsigma_1=\left( \begin{array}{ccc}
          \sigma_{11} & 0 & 0 \\
          0 & \sigma_{22} & 0 \\
          0 & 0 & \sigma_{33}
          \end{array}\right).
\end{equation}
The particles are embedded in a host  with a scalar conductivity $\sigma_2.$ In this paper, the principle axes of $\tsigma_1$ are assumed to be parallel to the geometric axes of the ellipsoidal particle. The orientation of the conductivity tensor  $\tsigma_1$  differs from inclusion to inclusion, such that $\tsigma_1$ is transformed to
\be
\tilde{\tsigma}_1 = \tnu\,\tsigma_1\,\tnu^T
\ee
 in a common coordinate system,  where $\tnu$ denotes the rotation matrix of a particle.
For the volume average of the field  of the whole system  $\langle\tE\rangle$   , one has~\cite{BergmanSSP92}
\be
p\langle\tE_1\rangle + (1-p) \langle\tE_L\rangle = \langle\tE\rangle,\label{AE0}
\ee
where $p$ stands for the volume fraction of particles, and $\langle\tE\rangle$ is equal to the applied field $\tE_0$  under appropriate boundary conditions~\cite{StroudPRB88}. Here $\langle\cdots\rangle$ denotes the volume average of $\cdots.$  $\langle\tE_L\rangle$ represents the volume average of the Lorentz field (namely, local field in the vicinity of the particles) which includes the contribution from the dipole moments of all the other particles, and $\langle\tE_1\rangle$ represents the volume average of the local field inside the particles.  Thus, we can rewrite Eq.~(\ref{AE0}) as~\cite{BergmanSSP92,StroudPRB88}
\be
p\langle\tE_1\rangle + (1-p) \langle\tE_L\rangle = \tE_0,\label{AE}
\ee
Solving the electrostatic equation $\nabla^2\Phi_1 = 0$ ($\Phi_1$ denotes the electrical potential inside the particles), we obtain  $\langle\tE_1\rangle$  such that~\cite{BergmanSSP92}
\be
\langle\tE_1\rangle = \langle\tilde{\trho}\rangle \langle\tE_L\rangle\label{AE1}
\ee
with
\be
\langle\tilde{\trho}\rangle =\langle \tnu \trho \tnu^T\rangle.
\ee
Here 
\be
\trho = \left( \begin{array}{ccc}
          \frac{\sigma_2}{L_x\sigma_{11}+(1-L_x)\sigma_{2}} & 0 & 0 \\
          0 &  \frac{\sigma_2}{L_y\sigma_{22}+(1-L_y)\sigma_{2}} & 0 \\
          0 & 0 &  \frac{\sigma_2}{L_z\sigma_{33}+(1-L_z)\sigma_{2}}
          \end{array}\right).
\ee

On the other hand, we average the electric current density $\langle\tJ\rangle$ over the volume of the whole system, and obtain
\be
\langle\tJ\rangle =  p\langle\tilde{\tsigma}_1\rangle\langle\tE_1\rangle + (1-p)\sigma_2 \langle\tE_L\rangle  .\label{AJ}
\ee
Further, the effective electrical conductivity $\tsigma_e$   of the ferrogel can be defined as
\be
\tsigma_e = \frac{\langle\tJ\rangle}{\langle\tE\rangle} = \frac{\langle \tJ\rangle}{\tE_0}.
\ee
In view of Eqs.~(\ref{AE}),~(\ref{AE1})~and~(\ref{AJ}),  we obtain
\be
\tsigma_e  = \frac{p\langle\tilde{\tsigma}_1\rangle\langle\tilde{\trho}\rangle+(1-p)\sigma_2\tI}{p\langle\tilde{\trho}\rangle+(1-p)\tI},\label{Eff}
\ee
which can be expressed in a tensorial form such that
\be
\tsigma_e = \left( \begin{array}{ccc}
         \sigma_{{\rm e11}} & 0 & 0 \\
          0 &  \sigma_{{\rm e22}} & 0 \\
          0 & 0 &  \sigma_{{\rm e33}}
          \end{array}\right).
\ee
In Eq.~(\ref{Eff}), $\tI$ represents a unit matrix.

Eq.~(\ref{Eff}) is a nontrivial equation, which can reproduce many other known results. For instance, setting $L_x=L_y\ne L_z$ and $\sigma_{11}=\sigma_{22}=\sigma_{33}$, and seeing the particles {\it randomly} oriented in a host, $\sigma_{{\rm e11}}$ obtained from Eq.~(\ref{Eff}) is identical to Eq.~(14) of Ref.~\cite{GaoJPCM}. In this case, if the metallic particles are subjected to an external magnetic field (i.e., no longer randomly oriented), $\sigma_{{\rm e11}}(=\sigma_{{\rm e22}})$ and $\sigma_{{\rm e33}}$ predicted by Eq.~(\ref{Eff}) are identical to Eqs.~(3)~and~(4) of Ref.~\cite{Rasa}, respectively. Finally, for spherical particles (namely, $L_x=L_y=L_z$) with a scalar conductivity (i.e., $\sigma_{11}=\sigma_{22}=\sigma_{33}$), Eq.~(\ref{Eff}) can produce the expression for the well-known Maxwell Garnett theory~\cite{MGA1,MGA2}, namely
\be
\sigma_e = \sigma_2+\frac{3p\sigma_2 (\sigma_1-\sigma_2)}{3\sigma_2+(1-p)(\sigma_1-\sigma_2)}.
\ee

\section{Numerical results}

We are now in a position to do some numerical calculations. We shall discuss two cases: (Case I)  Spherical particle with tensorial conductivity; (Case II) Ellipsoidal particle with scalar conductivity. In doing so, we can focus on the effect of shape or material (conductivity) anisotropies, respectively.

To start, we have to derive the expression for $\langle\tilde{\trho}\rangle$.  For $L_x=L_y\ne L_z$ and $\sigma_{11}=\sigma_{22}\ne \sigma_{33}$, $\langle\tilde{\trho}\rangle$ can be obtained as~\cite{Levy}
\be
\langle\tilde{\trho}\rangle = \left( \begin{array}{ccc}
         \frac{1}{2}[\rho_{11}+\rho_{33}-(\rho_{33}-\rho_{11})\zeta]  & 0 & 0 \\
          0 & \frac{1}{2}[\rho_{22}+\rho_{33}-(\rho_{33}-\rho_{22})\zeta ]  & 0 \\
          0 & 0 & \rho_{11}+(\rho_{33}-\rho_{11})\zeta  
          \end{array}\right),
\ee
where $\zeta = \langle\cos^2\theta\rangle, $ with  $\theta$ being the polar orientation angel of a particle, $0\le \theta\le\pi/2$. Since the orientation of the ferromagnetic particles in ferrogels can be affected by the external magnetic field, there is  $\zeta =1-2 L(\alpha)/\alpha$~\cite{Rasa}.  Here $L(\alpha)$ denotes the Langevin function and $\alpha=mH/k_BT$ the Langevin parameter, where $m$ represents the magnetic moment of a particle, and $H$ the magnetic field, $k_B$ the Boltzmann constant, and $T$ the temperature. For the sake of convenience, $\alpha$ will be used to denote the magnetic field in the following numerical calculations. In a word, $\zeta$ contains the information of the spatial orientation of the particles in the host. In particular, the zero field ($\alpha = 0$) yields  $\zeta \to 1/3$ naturally due to the random orientation of the particles. As $\alpha > 0,$ $\zeta$ deviates from $1/3$ accordingly. In other words, the specific spatial orientation  of the particles in the ferrogel can be achieved by adjusting the external magnetic field.

\subsection{Case I: Spherical particles with anisotropic conductivity}

In this case, there are $L_x=L_y=L_z=1/3$ and $\sigma_{11}=\sigma_{22}\ne \sigma_{33}$. In Fig.~1, we display the effective electrical conductivity versus the Langevin parameter for different conductivity ratio $t\, (=\sigma_{33}/\sigma_{11})$. In fact, $t$ represents the degree of anisotropy of conductivity.
 We find increasing the  Langevin parameter   $\alpha$ causes $\sigma_{{\rm e11}}$  to decrease, see Fig.~1(a)-(b).  In contrast, in Fig.~1(c)-(d), as $\alpha$ increases, $\sigma_{{\rm e33}}$ increases first, and decreases after a peak.  In particular,  at $\alpha=2.4$ there is always a peak for $\log_{10}[\sigma_{{\rm e33}}/\sigma_2]$. In detail, for $t=10^{\pm 12}$, $10^{\pm 8}$, and $10^{\pm 4}$, the corresponding peak values are $10.92$, $6.92$, and $2.92$, respectively. In other words,  an optimal magnetic field exists which leads to a maximum conductivity $\sigma_{{\rm e33}}$.
It is worth noting that both $t=10^{-N}$ and $t=10^N$ ($N$ is a number) can predict the same $\sigma_{{\rm e11}}$ or $\sigma_{{\rm e33}}$ concerned, see Fig.~1(b)~and~(d). This behavior keeps unchanged  up to $N = 19$ or so.  Moreover, larger $N$ and hence a stronger conductivity anisotropy can lead to a higher effective electrical conductivity  $\sigma_{{\rm e11}}$ or $\sigma_{{\rm e33}}$.

\subsection{Case II: Ellipsoidal particles with isotropic conductivity}

For this case,  $L_x=L_y\ne L_z$, and $\sigma_{11}=\sigma_{22}=\sigma_{33}$. Fig.~2 shows the effective electrical  conductivity  against the Langevin parameter $\alpha$ for different depolarization factor $L_z$.  For prolate spheroidal particles ($L_z<1/3$), smaller $L_z$ can lead to higher  $\sigma_{{\rm e11}}$ or $\sigma_{{\rm e33}}$, see Fig.~2. In contrast, for oblate spheroidal particles ($L_z>1/3$), larger $L_z$ can yield higher  $\sigma_{{\rm e11}}$ or lower $\sigma_{{\rm e33}}$. Also, for the prolate spheroidal cases, increasing the magnetic field causes  $\sigma_{{\rm e11}}$  (or $\sigma_{{\rm e33}}$) to decrease (or increase). However, for the oblate spheroidal cases, as the magnetic field increases,  $\sigma_{{\rm e11}}$  increases, but $\sigma_{{\rm e33}}$ decreases.  In a word, the stronger shape anisotropy (namely, how $L_z$ deviates from $1/3$) can yield a higher effective electrical  conductivity.

From Cases~I~and~II, it is concluded that  material (conductivity) anisotropies are more important to yield higher effective electrical  conductivities of ferrogels than shape anisotropies, while magnetic fields can offer a correction.

In addition, we also discussed the effect of the volume fraction of particles (no figures shown here). It is shown that increasing the volume fraction of particles can cause the effective electrical  conductivity to increase accordingly. This result is the same  as that obtained by Xue~\cite{Xue04}.

\section{Discussion and conclusion}

Here some comments are in order. In this paper, we have presented a general model for a ferrogel in which both the particle shape and conductivity have a tensorial form, in an attempt to study the effect of the shape and/or material (conductivity) anisotropy on the effective electrical  conductivity of the ferrogel, by taking into account the spatial orientation of the particles in the presence of an external magnetic field.
Our theory  reproduced several known results, and provides a link between the particle property and  orientation distribution and the effective electrical  conductivity of  ferrogels. 

For water, the minerals like sodium chloride are often dissolved in it, thus making it have a conductivity. More importantly,  as one tries to make pure water by gradually removing electrolytes, its conductivity gradually decreases indeed. However, if all electrolytes are removed, its conductivity is still nonzero. The reason is that an infinitesimal part of the molecules of water--only about one in 500 million--is ionized as hydrogen ions (H$^+$) and hydroxide ions (OH$^-$). Theoretically, at this point, the conductivity becomes $5.48\times 10^{-6}\,$S/m at $25\,$$^o$C. In contrast, the conductivity of polymers could be $\sim 10^{-15}\,$S/m. For the ferrogel discussed in this paper, we see the ferromagnetic particles to be embedded in the host medium which is composed of polymers as well as water.

Since the structure and magnetization property of a polydisperse ferrogel can differ from that of a monodisperse system~\cite{Ivanov,Huke,Wang}.  It is instructive to extend the present work to polydisperse case, so that one could investigate the polydisperse effect on the effective electrical  conductivity of polydisperse ferrogels.

In this work, we derived an anisotropic Maxwell-Garnett formula [Eq.~(\ref{Eff})], which is nonsymmetrical and may thus be suitable for low concentrations. For a higher concentration of particles, we can derive an alternative anisotropic Bruggeman formula (see Appendix),  which is symmetrical. Similar results should be obtained by means of the anisotropic Bruggeman formula. However, the Bruggeman formula can predict a percolation threshold.

It is also interesting to see what happens if one extend the present model to investigate the electro-optical effects~\cite{Levy},  nonlinearity enhancement~\cite{GaoPRB}, figure of merit~\cite{GaoPRB}, and nonlinear ac responses~\cite{Levy2,Hui98,HuangPRE01} of ferrogels due to the presence of the (weak) nonlinearity inside the metallic particles. In doing so, the particle reorientation arising from external magnetic fields is expected to play a role as well. In addition, due to the analogy  of the mathematical form between conductivities and dielectric constants of composite materials, it is straightforward to apply the present theory to the effective dielectric constant of ferrogels.

To sum up, based on our general model, it was shown that  material (conductivity) anisotropies are more important to produce a high effective electrical  conductivity of ferrogels than shape anisotropies, while magnetic fields can offer a  correction due to the change in orientation distributions of particles.

\section*{Acknowledgments}


This work was supported  by
the Deutsche Forschungsgemeinschaft (German Research Foundation)  under Grant No. HO 1108/8-3. I would like to thank Dr. C. Holm and Professor H. Pleiner for their critical comments, and   Professor K. W. Yu for  fruitful discussions.

\section*{Appendix: Anisotropic Bruggeman formula}

To derive the anisotropic Bruggeman formula, let us start from the effective dipole factor of the two components in the system. For the particles, the dipole factor $\tb_1$ is
\be
\tb_1 =  \left( \begin{array}{ccc}
          \frac{L_x(\sigma_{11}-\sigma_{{\rm e11}})}{L_x\sigma_{11}+(1-L_x)\sigma_{{\rm e11}}} & 0 & 0 \\
          0 &  \frac{L_y(\sigma_{22}-\sigma_{{\rm e22}})}{L_y\sigma_{22}+(1-L_y)\sigma_{{\rm e22}}} & 0 \\
          0 & 0 &  \frac{L_z(\sigma_{33}-\sigma_{{\rm e33}})}{L_z\sigma_{33}+(1-L_z)\sigma_{{\rm e33}}}
          \end{array}\right).
\ee
On the other hand, the host have a dipole factor $\tb_2$ such that
\be
\tb_2 = \frac{\sigma_2\tI-\tsigma_e}{\sigma_2\tI+2\tsigma_e}.
\ee 
Now we can derive the Bruggeman formula by considering the fact that the effective dipole factor $\langle\tb\rangle$ of the whole system should be zero, and then obtain
\be
\langle\tb\rangle\equiv  p\langle\tilde{\tb}_1\rangle + (1-p)\tb_2 = 0,\label{Bru}
\ee
where $\langle\tilde{\tb}_1\rangle = \langle\tnu \tb_1\tnu^T\rangle$.
In case of $L_x=L_y=L_z$ and $\sigma_{11}=\sigma_{22}=\sigma_{33}$ (scalar conductivity), Eq.~(\ref{Bru}) reduces to the well-know (isotropic) Bruggeman formula
\be
p\frac{\sigma_1-\sigma_e}{\sigma_1+2\sigma_e} + (1-p) \frac{\sigma_2-\sigma_e}{\sigma_2+2\sigma_e}=0.
\ee

 \newpage

\begin{figure}[h]
\caption{{\it Case I:} Effective electrical  conductivity, (a)-(b) $\sigma_{{\rm e11}}/\sigma_2$ and (c)-(d) $\sigma_{{\rm e33}}/\sigma_2$, versus Langevin parameter $\alpha ,$ for different conductivity ratio $t\,(=\sigma_{{\rm 33}}/\sigma_{{\rm 11}}).$ Parameters: $p=0.1$ and $\sigma_{11}/\sigma_2=10^{22}.$  }
\end{figure}

\begin{figure}[h]
\caption{{\it Case II:} Same as Fig.~1, but for different depolarization factor $L_z$.    }
\end{figure}

\newpage
\centerline{\epsfig{file=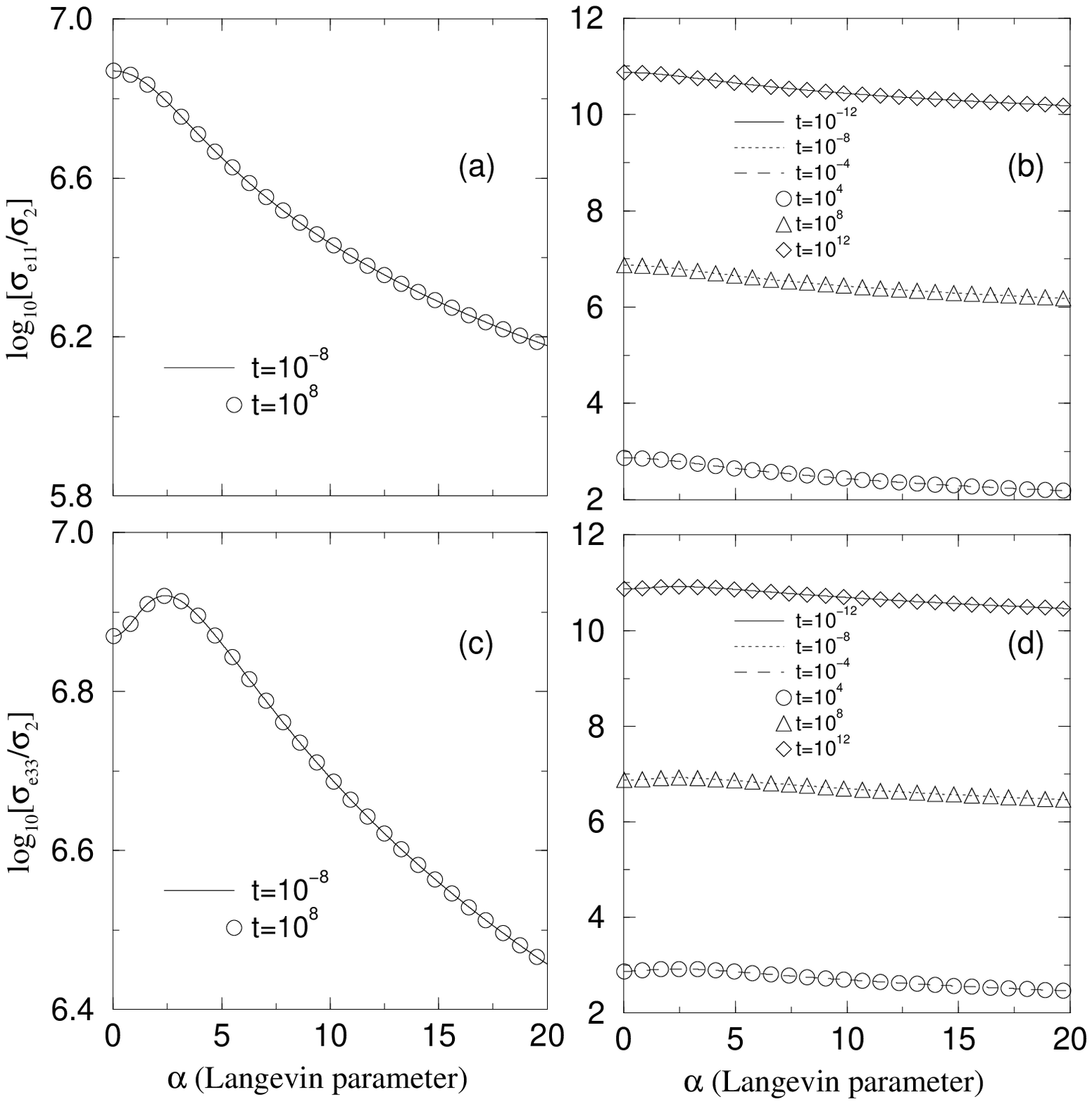,width=\linewidth}}
\centerline{Fig.~1}

\newpage
\centerline{\epsfig{file=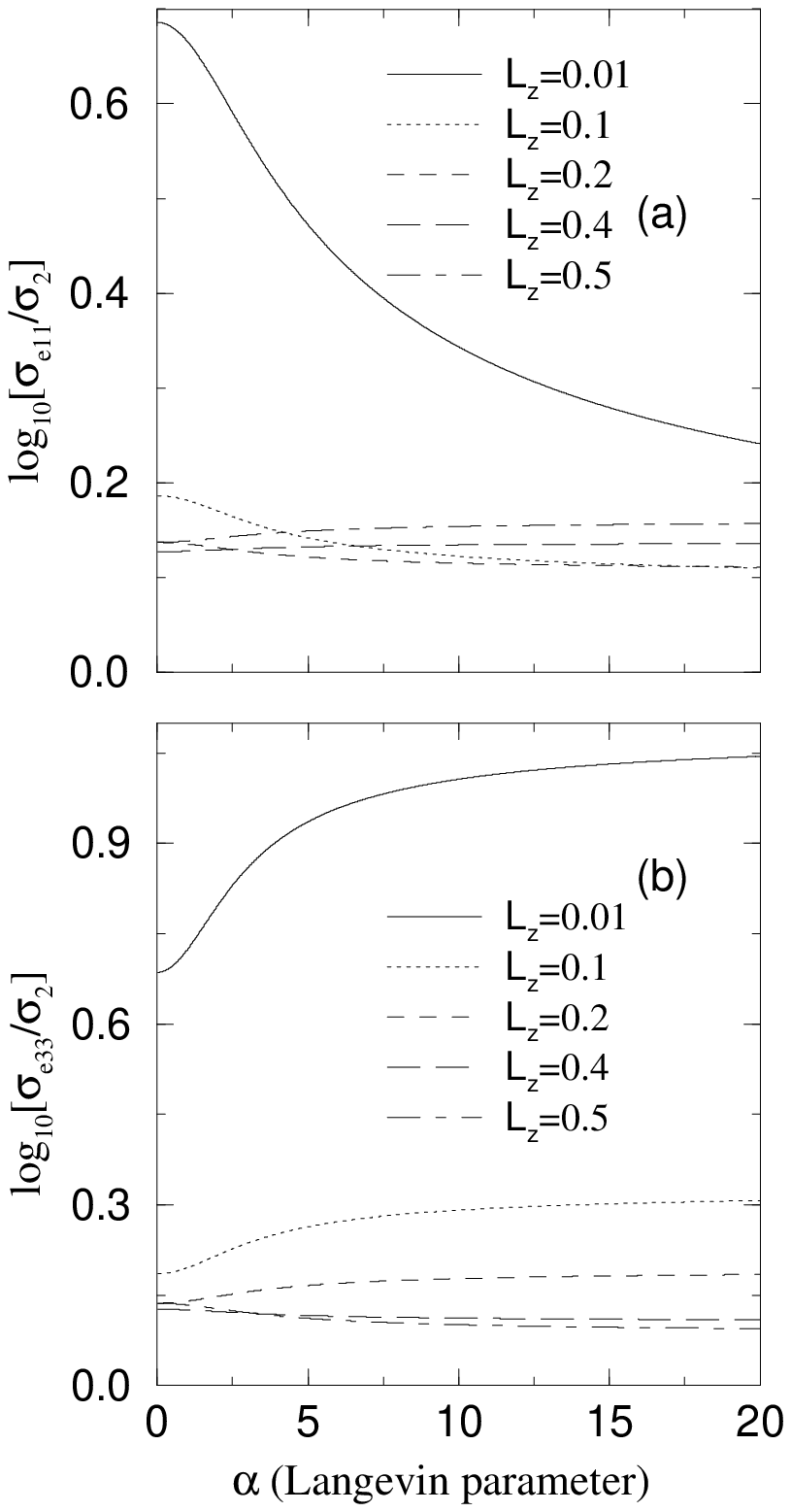,width=250pt}}
\centerline{Fig.~2}

\end{document}